\documentstyle[editedbook,epsfig,psfig,epsf]{mq}

\begin{opening}
\title{Short term X-ray rms variability of Cyg~X-1}
\author{T.~Gleissner$^1$, J.~Wilms$^1$, K.~Pottschmidt$^{2,3}$, P.~Uttley$^4$, M.A.~Nowak$^5$}
\author{R.~Staubert$^1$}
\institute{$^1$ Universit\"at T\"ubingen, IAAT -- Abt.~Astronomie, Sand 1, 72076 T\"ubingen, Germany\\
$^2$ Max-Planck-Institut f\"ur extraterr. Physik, Postfach 1312, 85748 Garching, Germany\\
$^3$ INTEGRAL Science Data Center, 16,\,Chemin d'\'Ecogia, 1290 Versoix, Switzerland\\
$^4$ Department of Physics and Astronomy, University of Southampton,
Southampton S017~1BJ, UK\\
$^5$ MIT-CXC, NE80-6077, 77 Massachusetts Ave., Cambridge, MA 02139, USA}
\end{opening}

\runningtitle{Short term X-ray rms variability of Cyg~X-1}
\runningauthor{Gleissner, Wilms, Pottschmidt, Uttley, Nowak \& Staubert}

\begin{document}
\vspace{-0.5cm}
\begin{abstract}
  {\small A linear dependence of the amplitude of broadband noise
    variability on flux for GBHC and AGN has been recently shown by Uttley
    \& McHardy \cite{uttley:01}. We present the long term evolution of this
    rms-flux-relation for Cyg~X-1 as monitored from 1998--2002 with RXTE.
    We confirm the linear relationship in the hard state and analyze the
    evolution of the correlation for the period of 1996--2002. In the
    intermediate and the soft state, we find considerable deviations from
    the otherwise linear relationship. A possible explanation for the
    rms-flux-relation is a superposition of local mass accretion rate
    variations.}
\end{abstract}

\section{Data Analysis}
The Cyg~X-1 data presented here were obtained by the PCA onboard RXTE,
mostly gained through our monitoring campaign 1998--2002 (for a description
of the data see \cite{pottschmidt:02a}). We split each light
curve into segments of 1\,s length and determine the mean flux of each
segment. The segments are binned into 41 equally segmented flux bins, for
each of which we calculate the power spectral density (PSD) of all
contained lightcurve segments via the DFT (see, e.g., \cite{{nowak:98b},{nowak:00a}}).

From all periodograms of each flux bin the mean PSD is determined using
standard methods. For better statistics we chose to take into account only
flux bins containing at least 20 periodograms.  Integrating the Poisson
noise corrected PSD over the range $\nu=1$--$32$\,Hz, we arrive at the squared
fractional rms variability. In a next step we multiply the fractional rms
variability by the mean flux of the bin, to obtain the absolute rms
variability $\sigma$.

\section{rms-flux-Relation}
For all bins of each observation we plot $\sigma$ over flux $F$ and fit a
linear function in two different representations. By fitting $\sigma = k F
+ a$, two characterizing values are determined: slope $k$ and intercept
$a$ on the $\sigma$ axis. The gradient of the $\sigma$-$F$-trend $k$ is
equivalent to the fractional rms variability of the light curve.

\begin{figure}[htb]
\centering 
\epsfig{file=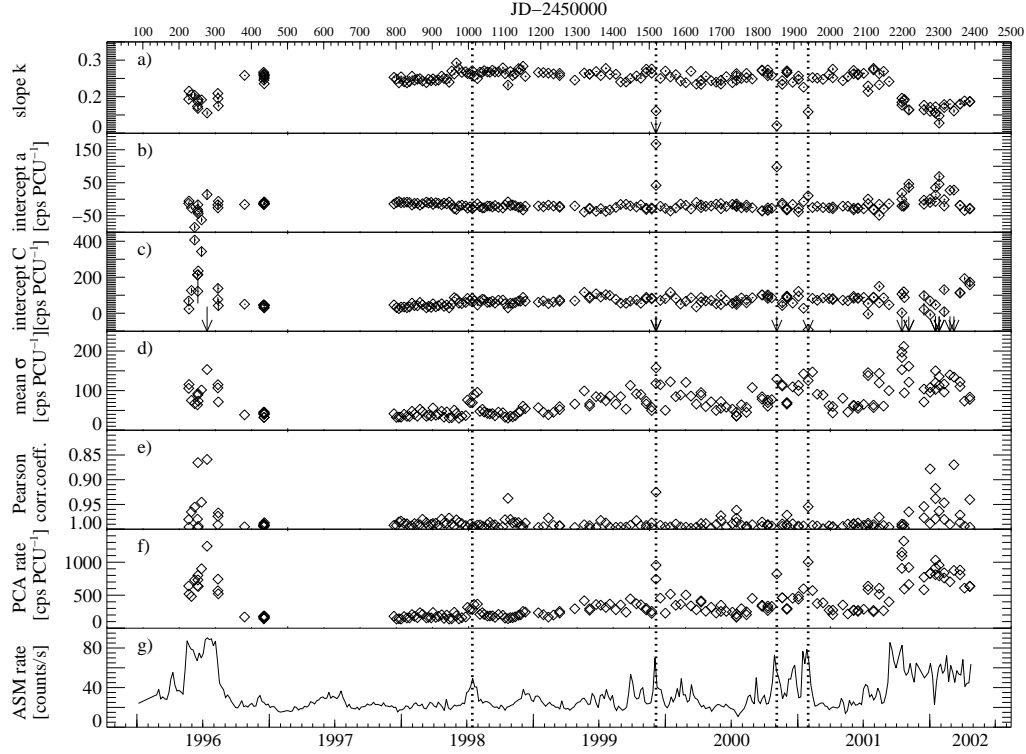,width=\columnwidth}
\caption{Evolution of fitting parameters and Cyg~X-1 system
  properties. 
  \textbf{a)} Fitting parameter slope $k$.
  \textbf{b)} Intercept $a$ on the $y$-axis.
  \textbf{c)} Intercept $C$ on the $x$-axis.
  \textbf{d)} Mean value of the absolute rms
  variability $\langle\sigma\rangle$.
  \textbf{e)} Pearson correlation coefficient for the correlation
  between $\sigma$ and flux $F$.
  \textbf{f)} Mean PCA count rate.
  \textbf{g)} Mean ASM count rate.}
\label{fig:rmsfluxvstime}
\end{figure}

Alternatively to intercept $a$, it was proposed to determine the physically
meaningful intercept $C$ on the $F$ axis by fitting $\sigma=k (F - C)$
\cite{uttley:01}. If $k$ is interpreted as the variable component of the
lightcurve, then $C$ represents a second component of it which does not
follow the linear $\sigma$-$F$-trend. In deriving results, it should be
kept in mind that $C$, which is identical to $-a/k$, and $k$ are not
independent values for the characterization of the rms-flux-relation.
Generally, there is a good linear relationship between $\sigma$ and $F$ in
the hard state, which is reflected by the stable Pearson correlation
coefficient in Fig.~\ref{fig:rmsfluxvstime}. We also notice coincidence of
the change of the general long term behavior of Cyg~X-1 from a ``quiet hard
state'' to a ``flaring hard state'' in 1998 May \cite{pottschmidt:02a} with
a shift in the values of slope $k$ and the intercepts.

\section{In the Soft State and during ``Failed State Transitions''}
\begin{figure}
\centering 
\epsfig{file=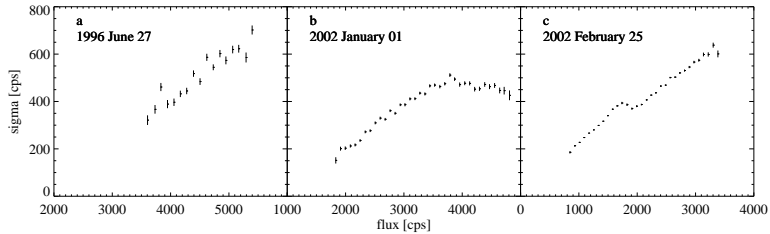,width=10cm}
\caption{Exemplary soft state shapes: \textbf{a)} linearly cluttered, \textbf{b)} arch-like and
  \textbf{c)} wavy.}
\label{fig:softstate}
\end{figure}

The shape of the rms-flux-relation in the soft states of 1996 and 2001/2002
changes continuously between waviness and approximate linearity (see
Fig.~\ref{fig:softstate}). 

\begin{figure}
\centering 
\epsfig{file=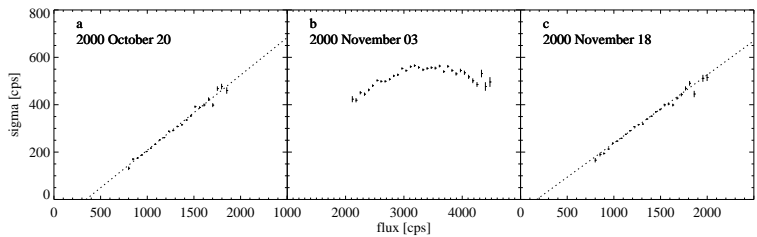,width=10cm}
\caption{Failed state transition (FST) 2000 Nov 03: the
  observations \textbf{a)} two weeks before and \textbf{c)} two weeks after
  show the common linear relation, while the relation in the FST \textbf{b)}
  is arch-like. (dotted lines: linear fitting function)}
\label{fig:failed}
\end{figure}

During ``failed state transitions''(FST), i.e., during times when the
source reached its intermediate state
\cite{{pottschmidt:02a},{belloni:96a}}, the linear rms-flux relation breaks
down. We examined four FST: 1998 July 15, 1999 Dec 05, 2000 Nov 03, and
2001~Jan~29 (indicated in Fig.~\ref{fig:rmsfluxvstime} by dotted lines).
Comparing these events with the neighboring hard state observations, we
notice that the rms-flux-relations of 1999 December~05 and 2000 November 03
-- being strictly linear before and after -- change to an arch-like
appearance during the FST itself (see Fig.~\ref{fig:failed}). The other two
examined FST also deviate from the linear relation, but these deviations are
not as prominent.

\section{Discussion and Conclusions}
An explanation of the rms-flux-relation seems to be the modification of a
theory which explains the PSD of GBHC by the superposition of accretion
disk (AD) instabilities occuring at several radii in the AD that are
propagated towards the innermost region of the disk
\cite{{lyubarskii:97a},{churazov:01a}}. Local $\dot{M}$ variations, which
are caused by short term changes of the disk viscosity and whose time scale
is dependent on radius, are superimposed on long term $\dot{M}$
modulations. If the fractional amplitude of the $\dot{M}$ variations is
independent of the long term $\dot{M}$, a linear flux-rms relation will be
observed \cite{uttley:01}. The rms-flux-relation will break down if the
corona is severely disturbed. This seems to be the case during the
intermediate state of Cyg~X-1, in agreement with earlier results for the
behavior of X-ray time lags \cite{pottschmidt:00a}. Here, the observed
X-ray time lag is much larger than during the normal hard state, which
could indicate changes in the geometry of the AD corona, possibly related
to the observed radio emitting outflow.

\end{document}